\documentclass[12 pt,twoside,aps,prd,amsmath,amssymb,
tightenlines,eqsecnum,showpacs,showkeys]{revtex4}
\usepackage{mathptmx}
\usepackage{bm}
\usepackage{graphicx}
\usepackage{hyperref}
\renewcommand{\cal}{\mathcal}

\newcommand {\ve}{\varepsilon}
\newcommand {\pr}{\partial}
\newcommand {\prm}{\prime}
\newcommand {\cG}{\cal G}
\newcommand {\cD}{\cal D}
\newcommand {\G}{\Gamma}
\newcommand {\bg}{\bar \gamma}
\newcommand {\bp}{\bar \psi}

\newcommand {\p}{\psi}

\def \myfigures #1#2#3#4#5#6#7#8
{\begin{figure}[ht]
    \begin{center}
        \includegraphics[width=#2 \textwidth]{#1.eps}
        \hfill
        \includegraphics[width=#6 \textwidth]{#5.eps}
        \parbox[t]{#4\textwidth}{\caption {#3}\label{#1}}
        \hfill
        \parbox[t]{#8\textwidth}{\caption {#7}\label{#5}}
    \end{center}
\end{figure} }

\def\myfigure #1#2#3#4
{\begin{figure}[ht]\begin{center}
\includegraphics[width=#2 \textwidth]{#1.eps}
\parbox[t]{#4\textwidth}{\caption{#3}\label{#1}}
\end{center}\end{figure}}

\date{\today}
\begin{document}
\title{Nonlinear spinor field in Bianchi type-I Universe filled
with viscous fluid: numerical solutions}
\author{Bijan Saha}
\affiliation{Laboratory of Information Technologies\\
Joint Institute for Nuclear Research, Dubna\\
141980 Dubna, Moscow region, Russia} \email{bijan@jinr.ru} \homepage{http://www.jinr.ru/~bijan/}

\begin{abstract}
We consider a system of nonlinear spinor and a Bianchi type I
gravitational fields in presence of viscous fluid. The nonlinear
term in the spinor field Lagrangian is chosen to be $\lambda F$,
with $\lambda$ being a self-coupling constant and $F$ being a
function of the invariants $I$ an $J$ constructed from bilinear
spinor forms $S$ and $P$. Self-consistent solutions to the
spinor and BI gravitational field equations are obtained in terms
of $\tau$, where $\tau$ is the volume scale of BI universe. System
of equations for $\tau$ and $\ve$, where $\ve$ is the energy of
the viscous fluid, is deduced. This system is solved numerically for some
special cases.

\end{abstract}



\maketitle

\bigskip


\section{Introduction}

The investigation of relativistic cosmological models usually has
the energy momentum tensor of matter generated by a perfect fluid.
To consider more realistic models, one must take into account the
viscosity mechanisms, which have already attracted attention
of many researchers. Misner \cite{mis1,mis2} suggested that strong
dissipative due to the neutrino viscosity may considerably reduce
the anisotropy of the black-body radiation. Viscosity mechanism in
cosmology can explain the anomalously high entropy per baryon in
the present universe \cite{wein,weinb}. Bulk viscosity associated
with the grand-unified-theory phase transition \cite{lang} may
lead to an inflationary scenario \cite{waga,pacher,guth}.

A uniform cosmological model filled with fluid which possesses pressure
and second (bulk) viscosity was developed by Murphy \cite{murphy}. The
solutions that he found exhibit an interesting feature that the big bang
type singularity appears in the infinite past. Exact solutions of the
isotropic homogeneous cosmology for open, closed and flat universe have
been found by Santos et al \cite{santos}, with the bulk viscosity being
a power function of energy density.

The nature of cosmological solutions for homogeneous Bianchi type
I (BI) model was investigated by Belinsky and Khalatnikov
\cite{belin} by taking into account a dissipative process due to
viscosity. They showed that viscosity cannot remove the
cosmological singularity but results in a qualitatively new
behavior of the solutions near singularity. They found the
remarkable property that during the time of the \textit{big bang}
matter is created by the gravitational field. BI solutions in case
of stiff matter with a shear viscosity being the power function of
of energy density were obtained by Banerjee \cite{baner}, whereas
BI models with bulk viscosity ($\eta$) that is a power function of
energy density $\ve$ and when the universe is filled with stiff
matter were studied by Huang \cite{huang}. The effect of bulk
viscosity, with a time varying bulk viscous coefficient, on the
evolution of isotropic FRW models was investigated in the context
of open thermodynamics system was studied by Desikan
\cite{desikan}. This study was further developed by Krori and
Mukherjee \cite{krori} for anisotropic Bianchi models.
Cosmological solutions with nonlinear bulk viscosity were obtained
in \cite{chim}. Models with both shear and bulk viscosity were
investigated in \cite{elst,meln}.

Though Murphy \cite{murphy} claimed that the introduction of bulk
viscosity can avoid the initial singularity at finite past,
results obtained in \cite{barrow} show that, it is, in general,
not valid, since for some cases big bang singularity occurs in
finite past. To eliminate the initial singularities a
self-consistent system of nonlinear spinor and BI gravitational
field was considered by us in a series of papers
\cite{sahajmp,sahaprd,sahagrg,sahal}. For some cases we were able
to find field (both matter and gravitational) configurations those
were always regular. In the papers mentioned above we considered
the system of interacting nonlinear spinor and/or scalar fields in
a BI universe filled with perfect fluid. We also study the above
system in presence of cosmological constant $\Lambda$ (both
constant and time varying \cite{sahal}). A nonlinear spinor field,
suggested by the symmetric coupling between nucleons, muons, and
leptons, has been investigated by Finkelstein et. al.
\cite{finkel} in the classical approximation. Although the
existence of spin-$1/2$ fermion is both theoretically and
experimentally undisputed, these are described by {\it quantum}
spinor fields. Possible justifications for the existence of
classical spinors has been addressed in \cite{greene}. In view of
what has been mentioned above, it would be interesting to study
the influence of viscous fluid to the system of material (say
spinor and/or scalar) and BI gravitational fields in presence of a
$\Lambda$-term as well. In a recent paper we studied the Bianchi type-I
universe filled with viscous fluid in presence of a $\Lambda$ term
\cite{Visprd04}. This study was further developed in \cite{Visprd04} where
we present qualitative analysis of the corresponding system of equations.
Finally in \cite{Visprd04} we introduced spinor field into the system and
solved the system for some special choice of viscosity. The purpose of
this paper is to further developed those results for more general cases and
give some numerical results. It should be noted the in the process there occurs
a very rich system of equations for volume scale, Hubble constant and energy density.
The qualitative analysis of this system is under active study and we plan to present
those results soon.

        \section{Derivation of Basic Equations}
In this section we derive the fundamental equations for the
interacting spinor, scalar and gravitational fields from the
action and write their solutions in term of the volume scale
$\tau$ defined bellow \eqref{taudef}. We also derive the equation
for $\tau$ which plays the central role here.


We consider a system of nonlinear spinor, scalar and BI gravitational
field in presence of perfect fluid given by the action
\begin{equation}
{\cal S}(g; \p, \bp) = \int\, {\cal L} \sqrt{-g} d\Omega
\label{action}
\end{equation}
with
\begin{equation}
{\cal L} = {\cal L}_{\rm g} + {\cal L}_{\rm sp} + {\cal L}_{\rm
m}. \label{lag}
\end{equation}
The gravitational part of the Lagrangian \eqref{lag} is given by a
Bianchi type I (BI hereafter) space-time, whereas ${\cal L}_{\rm
sp}$ describes the spinor field lagrangian and ${\cal L}_{\rm m}$
stands for the lagrangian density of viscous fluid.

             \subsection{Material field Lagrangian}
For a spinor field $\p$, symmetry between $\p$ and $\bp$ appears
to demand that one should choose the symmetrized Lagrangian~
\cite{kibble}. Keep it in mind we choose the spinor field
Lagrangian as
\begin{equation}
{\cal L}_{\rm sp}=\frac{i}{2} \biggl[ \bp \gamma^{\mu}
\nabla_{\mu} \p- \nabla_{\mu} \bp \gamma^{\mu} \p \biggr] - m\bp
\p + \lambda F,  \label{nlspin}
\end{equation}
Here $m$ is the spinor mass, $\lambda$ is the self-coupling
constant and $F = F(I,J)$ with $I = S^ 2= (\bar \psi \psi)^2$ and
$J = P^2 = (i \bar \psi \gamma^5 \psi)^2$. According to the
Pauli-Fierz theorem \cite{Ber} among the five invariants only $I$
and $J$ are independent as all other can be expressed by them:
$I_V = - I_A = I + J$ and $I_Q = I - J.$ Therefore, the choice $F
= F(I, J)$, describes the nonlinearity in the most general of its
form \cite{sahaprd}. Note that setting $\lambda = 0$ in
\eqref{nlspin} we come to the case with linear spinor field.

             \subsection{The gravitational field}
As a gravitational field we consider the Bianchi type I (BI) cosmological
model. It is the simplest model of anisotropic universe that describes
a homogeneous and spatially flat space-time and if filled with perfect
fluid with the equation of state $p = \zeta \ve, \quad \zeta < 1$, it
eventually evolves into a FRW universe \cite{jacobs}. The isotropy of
present-day universe makes BI model a prime candidate for studying the
possible effects of an anisotropy in the early universe on modern-day
data observations. In view of what has been mentioned above we choose
the gravitational part of the Lagrangian \eqref{lag} in the form
\begin{equation}
{\cal L}_{\rm g} = \frac{R}{2\kappa},
\label{lgrav}
\end{equation}
where $R$ is the scalar curvature, $\kappa = 8 \pi G$
being the Einstein's gravitational constant. The gravitational field in
our case is given by a Bianchi type I (BI) metric
\begin{equation}
ds^2 = dt^2 - a^2 dx^2 - b^2 dy^2 - c^2 dz^2,
\label{BI}
\end{equation}
with $a,\, b,\, c$ being the functions of time $t$ only. Here the speed of
light is taken to be unity.

  \subsection{Field equations}
Let us now write the field equations corresponding to the action
\eqref{action}.

Variation of \eqref{action} with respect to spinor field $\psi\,(\bp)$
gives spinor field equations
\begin{subequations}
\label{speq}
\begin{eqnarray}
i\gamma^\mu \nabla_\mu \psi - m \psi + {\cD} \psi +
{\cG} i \gamma^5 \psi &=&0, \label{speq1} \\
i \nabla_\mu \bp \gamma^\mu +  m \bp - {\cD} \bp -
{\cG} i \bp \gamma^5 &=& 0, \label{speq2}
\end{eqnarray}
\end{subequations}
where we denote
$$ {\cD} =  2\lambda S \frac{\pr F}{\pr I},
\quad
{\cG} = 2 \lambda P \frac{\pr F}{\pr J}.$$

Variation of \eqref{action} with respect to metric tensor
$g_{\mu\nu}$ gives the Einstein's equations which in account of
the $\Lambda$-term for the BI space-time \eqref{BI} can be
rewritten as
\begin{subequations}
\label{BID}
\begin{eqnarray}
\frac{\ddot b}{b} +\frac{\ddot c}{c} + \frac{\dot b}{b}\frac{\dot
c}{c}&=&  \kappa T_{1}^{1} +\Lambda,\label{11}\\
\frac{\ddot c}{c} +\frac{\ddot a}{a} + \frac{\dot c}{c}\frac{\dot
a}{a}&=&  \kappa T_{2}^{2} + \Lambda,\label{22}\\
\frac{\ddot a}{a} +\frac{\ddot b}{b} + \frac{\dot a}{a}\frac{\dot
b}{b}&=&  \kappa T_{3}^{3} + \Lambda,\label{33}\\
\frac{\dot a}{a}\frac{\dot b}{b} +\frac{\dot b}{b}\frac{\dot c}{c}
+\frac{\dot c}{c}\frac{\dot a}{a}&=&  \kappa T_{0}^{0} + \Lambda,
\label{00}
\end{eqnarray}
\end{subequations}
where over dot means differentiation with respect to $t$
and $T_{\nu}^{\mu}$ is the energy-momentum tensor
of the material field given by
\begin{equation}
T_{\mu}^{\nu} = T_{{\rm sp}\,\mu}^{\,\,\,\nu} + T_{{\rm m}\,\mu}^{\,\,\,\nu}.
\label{tem}
\end{equation}
Here $T_{{\rm sp}\,\mu}^{\,\,\,\nu}$ is the energy-momentum tensor of
the spinor field which with regard to \eqref{speq} has the form
\begin{eqnarray}
T_{{\rm sp}\,\mu}^{\,\,\,\rho}&=&\frac{i}{4} g^{\rho\nu}\biggl(\bp \gamma_\mu
\nabla_\nu \psi + \bp \gamma_\nu \nabla_\mu \psi - \nabla_\mu \bar
\psi \gamma_\nu \psi - \nabla_\nu \bp \gamma_\mu \psi \biggr)\label{temsp} \\
&+&\delta_{\mu}^{\rho} \bigl({\cD} S + {\cG} P - \lambda F\bigr).
\nonumber
\end{eqnarray}

$T_{{\rm m}\mu}^{\nu}$ is the energy-momentum tensor of a viscous
fluid having the form
\begin{equation}
T_{{\rm m}\mu}^{\nu} = (\ve + p^\prm) u_\mu u^\nu - p^{\prm}
\delta_\mu^\nu + \eta g^{\nu \beta} [u_{\mu;\beta}+u_{\beta:\mu}
-u_\mu u^\alpha u_{\beta;\alpha} - u_\beta u^\alpha
u_{\mu;\alpha}], \label{imper}
\end{equation}
where
\begin{equation}
p^{\prm} = p - (\xi - \frac{2}{3} \eta) u^\mu_{;\mu}.
\label{ppr}
\end{equation}
Here $\ve$ is the energy density, $p$ - pressure, $\eta$ and $\xi$
are the coefficients of shear and bulk viscosity, respectively. In
a comoving system of reference such that $u^\mu = (1,\,0,\,0,\,0)$
we have
\begin{subequations}
\begin{eqnarray}
T_{{\rm m}0}^{0} &=& \ve, \\
T_{{\rm m}1}^{1} &=& - p^{\prm} + 2 \eta \frac{\dot a}{a}, \\
T_{{\rm m}2}^{2} &=& - p^{\prm} + 2 \eta \frac{\dot b}{b}, \\
T_{{\rm m}3}^{3} &=& - p^{\prm} + 2 \eta \frac{\dot c}{c}.
\end{eqnarray}
\end{subequations}

In the Eqs. \eqref{speq} and \eqref{temsp} $\nabla_\mu$ is the covariant
derivatives acting on a spinor field as ~\cite{zhelnorovich,brill}
\begin{equation}
\label{cvd}
\nabla_\mu \psi = \frac{\partial \psi}{\partial x^\mu} -\G_\mu \psi, \quad
\nabla_\mu \bp = \frac{\partial \bp}{\partial x^\mu} + \bp \G_\mu,
\end{equation}
where $\G_\mu$ are the Fock-Ivanenko spinor connection coefficients
defined by
\begin{equation}
\G_\mu = \frac{1}{4} \gamma^\sigma \Bigl(\G_{\mu \sigma}^{\nu} \gamma_{\nu}
- \partial_{\mu} \gamma_{\sigma}\Bigr).
\label{fock}
\end{equation}
For the metric \eqref{BI} one has the following components
of the spinor connection coefficients
\begin{eqnarray}
\G_0 = 0, \quad
\G_1 = \frac{1}{2}\dot a(t) \bg^1 \bg^0, \quad
\G_2 = \frac{1}{2}\dot b(t) \bg^2 \bg^0, \quad
\G_3 = \frac{1}{2}\dot c(t) \bg^3 \bg^0.
\label{ficc}
\end{eqnarray}
The Dirac matrices $\gamma^\mu(x)$ of curved space-time are
connected with those of Minkowski one as follows:
$$ \gamma^0=\bg^0,\quad \gamma^1 =\bg^1/a,
\quad \gamma^2=\bg^2 /b,\quad \gamma^3 =\bg^3 /c$$
with
\begin{eqnarray}
\bg^0\,=\,\left(\begin{array}{cc}I&0\\0&-I\end{array}\right), \quad
\bg^i\,=\,\left(\begin{array}{cc}0&\sigma^i\\
-\sigma^i&0\end{array}\right), \quad
\gamma^5 = \bg^5&=&\left(\begin{array}{cc}0&-I\\
-I&0\end{array}\right),\nonumber
\end{eqnarray}
where $\sigma_i$ are the Pauli matrices:
\begin{eqnarray}
\sigma^1\,=\,\left(\begin{array}{cc}0&1\\1&0\end{array}\right),
\quad
\sigma^2\,=\,\left(\begin{array}{cc}0&-i\\i&0\end{array}\right),
\quad
\sigma^3\,=\,\left(\begin{array}{cc}1&0\\0&-1\end{array}\right).
\nonumber
\end{eqnarray}
Note that the $\bg$ and the $\sigma$ matrices obey the following
properties:
\begin{eqnarray}
\bg^i \bg^j + \bg^j \bg^i = 2 \eta^{ij},\quad i,j = 0,1,2,3
\nonumber\\
\bg^i \bg^5 + \bg^5 \bg^i = 0, \quad (\bg^5)^2 = I,
\quad i=0,1,2,3 \nonumber\\
\sigma^j \sigma^k = \delta_{jk} + i \varepsilon_{jkl} \sigma^l,
\quad j,k,l = 1,2,3 \nonumber
\end{eqnarray}
where $\eta_{ij} = \{1,-1,-1,-1\}$ is the diagonal matrix,
$\delta_{jk}$ is the Kronekar symbol and $\varepsilon_{jkl}$
is the totally antisymmetric matrix with $\varepsilon_{123} = +1$.

We study the space-independent solutions to the spinor
field equations \eqref{speq} so that
$\psi=\psi(t)$.
Here we define
\begin{equation}
\tau = a b c = \sqrt{-g}
\label{taudef}
\end{equation}

The spinor field equation \eqref{speq1} in account of \eqref{cvd} and
\eqref{ficc} takes the form
\begin{equation} i\bg^0
\biggl(\frac{\partial}{\partial t} +\frac{\dot \tau}{2 \tau} \biggr) \psi
- m \psi + {\cD}\psi + {\cG} i \gamma^5 \psi = 0.
\label{spq}
\end{equation}
Setting $V_j(t) = \sqrt{\tau} \psi_j(t), \quad j=1,2,3,4,$ from
\eqref{spq} one deduces the following system of equations:
\begin{subequations}
\label{V}
\begin{eqnarray}
\dot{V}_{1} + i (m - {\cD}) V_{1} - {\cG} V_{3} &=& 0, \\
\dot{V}_{2} + i (m - {\cD}) V_{2} - {\cG} V_{4} &=& 0, \\
\dot{V}_{3} - i (m - {\cD}) V_{3} + {\cG} V_{1} &=& 0, \\
\dot{V}_{4} - i (m - {\cD}) V_{4} + {\cG} V_{2} &=& 0.
\end{eqnarray}
\end{subequations}

From \eqref{speq1} we also write the equations for the invariants
$S,\quad P$ and $A = \bp \bg^5 \bg^0 \psi$
\begin{subequations}
\label{inv}
\begin{eqnarray}
{\dot S_0} - 2 {\cG}\, A_0 &=& 0, \label{S0}\\
{\dot P_0} - 2 (m - {\cD})\, A_0 &=& 0, \label{P0}\\
{\dot A_0} + 2 (m - {\cD})\, P_0 + 2 {\cG} S_0 &=& 0, \label{A0}
\end{eqnarray}
\end{subequations}
where $S_0 = \tau S, \quad P_0 = \tau P$, and $ A_0 = \tau A.$ The
Eq. \eqref{inv} leads to the following relation
\begin{equation}
S^2 + P^2 + A^2 =  C^2/ \tau^2, \qquad C^2 = {\rm const.}
\label{inv1}
\end{equation}

Giving the concrete form of $F$ from \eqref{V} one writes
the components of the spinor functions in explicitly and
using the solutions obtained one can write the components of
spinor current:
\begin{equation}
j^\mu = \bp \gamma^\mu \psi.
\label{spincur}
\end{equation}
The component $j^0$
\begin{equation}
j^0 = \frac{1}{\tau}
\bigl[V_{1}^{*} V_{1} + V_{2}^{*} V_{2} + V_{3}^{*} V_{3}
+ V_{4}^{*} V_{4}\bigr],
\end{equation}
defines the charge density of spinor field
that has the following chronometric-invariant form
\begin{equation}
\varrho = (j_0\cdot j^0)^{1/2}.
\label{rho}
\end{equation}
The total charge of spinor field is defined as
\begin{equation}
Q = \int\limits_{-\infty}^{\infty} \varrho \sqrt{-^3 g} dx dy dz =
   \varrho \tau {\cal V},
\label{charge}
\end{equation}
where ${\cal V}$ is the volume. From the spin tensor
\begin{equation}
S^{\mu\nu,\epsilon} = \frac{1}{4}\bp \bigl\{\gamma^\epsilon
\sigma^{\mu\nu}+\sigma^{\mu\nu}\gamma^\epsilon\bigr\} \psi.
\label{spin}
\end{equation}
one finds chronometric invariant spin tensor
\begin{equation}
S_{{\rm ch}}^{ij,0} = \bigl(S_{ij,0} S^{ij,0}\bigr)^{1/2},
\label{chij}
\end{equation}
and the projection of the spin vector on $k$ axis
\begin{equation}
S_k = \int\limits_{-\infty}^{\infty} S_{{\rm ch}}^{ij,0}
\sqrt{-^3 g} dx dy dz = S_{{\rm ch}}^{ij,0} \tau V.
\label{proj}
\end{equation}

Let us now solve the Einstein equations. To do it, we first write the
expressions for the components of the energy-momentum tensor explicitly:
\begin{subequations}
\label{total}
\begin{eqnarray}
T_{0}^{0} &=& m S -\lambda F + \ve \equiv \tilde{T}_{0}^{0},\\
T_{1}^{1} &=& {\cD} S + {\cG} P - \lambda F
- p^{\prm} + 2 \eta \frac{\dot a}{a} \equiv \tilde{T}_{1}^{1} + 2 \eta \frac{\dot a}{a}, \\
T_{2}^{2} &=& {\cD} S + {\cG} P - \lambda F
- p^{\prm} + 2 \eta \frac{\dot b}{b} \equiv \tilde{T}_{1}^{1} + 2 \eta \frac{\dot b}{b},, \\
T_{3}^{3} &=& {\cD} S + {\cG} P - \lambda F - p^{\prm} + 2 \eta
\frac{\dot c}{c} \equiv \tilde{T}_{1}^{1} + 2 \eta \frac{\dot
c}{c},.
\end{eqnarray}
\end{subequations}
In account of \eqref{total} subtracting \eqref{11} from \eqref{22},
one finds the following relation between $a$ and $b$:
\begin{equation}
\frac{a}{b}= D_1 \exp \biggl(X_1 \int \frac{
e^{-2 \kappa \int \eta dt}dt}{\tau}\biggr).
\label{ab}
\end{equation}
Analogically, one finds
\begin{eqnarray}
\frac{b}{c}= D_2 \exp \biggl(X_2 \int \frac{
e^{-2 \kappa \int \eta dt}dt}{\tau}\biggr), \quad
\frac{c}{a}= D_3 \exp \biggl(X_3 \int \frac{
e^{-2 \kappa \int \eta dt}dt}{\tau}\biggr).
\label{ac}
\end{eqnarray}
Here $D_1,\,D_2,\,D_3,\,X_1,\, X_2, X_3 $ are integration constants, obeying
\begin{eqnarray}
D_1 D_2 D_3 = 1, \quad X_1 + X_2 + X_3 = 0.
\label{intcon}
\end{eqnarray}

In view of \eqref{intcon} from \eqref{ab} and \eqref{ac}
we write the metric functions explicitly~\cite{sahaprd}
\begin{subequations}
\label{abc}
\begin{eqnarray}
a(t) &=&
(D_{1}/D_{3})^{1/3}\tau^{1/3}\exp \biggl[\frac{X_1 - X_3
}{3} \int\,\frac{e^{-2 \kappa \int \eta dt}}{\tau (t)} dt \biggr],
\label{a} \\
b(t) &=&
(D_{1}^{2}D_{3})^{-1/3}\tau^{1/3}\exp \biggl[-\frac{2 X_1 + X_3
}{3} \int\,\frac{e^{-2 \kappa \int \eta dt} }{\tau (t)} dt \biggr],
\label{b} \\
c(t) &=&
(D_{1}D_{3}^{2})^{1/3}\tau^{1/3}\exp \biggl[\frac{X_1 + 2 X_3
}{3} \int\,\frac{e^{-2 \kappa \int \eta dt}}{\tau (t)} dt \biggr].
\label{c}
\end{eqnarray}
\end{subequations}
As one sees from \eqref{a}, \eqref{b} and \eqref{c}, for $\tau = t^n$
with $n > 1$ the exponent tends to unity at large $t$, and the
anisotropic model becomes isotropic one.

Further we will investigate the existence of singularity (singular
point) of the gravitational case, which can be done by
investigating the invariant characteristics of the space-time. In
general relativity these invariants are composed from the
curvature tensor and the metric one. In a 4D Riemann space-time
there are 14 independent invariants. Instead of analyzing all 14
invariants, one can confine this study only in 3, namely the
scalar curvature $I_1 = R$, $I_2 = R_{\mu\nu}^R{\mu\nu}$, and the
Kretschmann scalar $I_3 =
R_{\alpha\beta\mu\nu}R^{\alpha\beta\mu\nu}$. At any regular
space-time point, these three invariants $I_1,\,I_2,\,I_3$ should
be finite. One can easily verify that
$$I_1 \propto \frac{1}{\tau^2},\quad
I_2 \propto \frac{1}{\tau^4},\quad I_3 \propto \frac{1}{\tau^4}.$$
Thus we see that at any space-time point, where $\tau = 0$ the invariants
$I_1,\,I_2,\,I_3$, as well as the scalar and spinor fields
become infinity, hence the space-time becomes singular at this point.

In what follows, we write the equation for $\tau$ and study it in details.

Summation of Einstein equations \eqref{11}, \eqref{22}, \eqref{33} and
\eqref{00} multiplied by 3 gives
\begin{equation}
\ddot \tau = \frac{3}{2}\kappa \Bigl(\tilde{T}_{0}^{0} +
\tilde{T}_{1}^{1}\Bigr)\tau + 3 \kappa \eta \dot \tau + 3 \Lambda
\tau, \label{a4}
\end{equation}
which can be rearranged as
\begin{equation}
{\ddot \tau} - \frac{3}{2} \kappa \xi {\dot \tau} =
\frac{3}{2}\kappa \Bigl(mS + {\cD} S + {\cG} P - 2 \lambda F + \ve
- p\Bigr) \tau + 3 \Lambda \tau. \label{dtau1}
\end{equation}
For the right-hand-side of \eqref{dtau1} to be a function
of $\tau$ only, the solution to this equation is well-known~\cite{kamke}.

On the other hand from Bianchi identity $G_{\mu;\nu}^{\nu} = 0$ one finds
\begin{equation}
T_{\mu;\nu}^{\nu} = T_{\mu,\nu}^{\nu} + \G_{\rho\nu}^{\nu} T_{\mu}^{\rho}
- \G_{\mu\nu}^{\rho} T_{\rho}^{\nu} = 0,
\end{equation}
which in our case has the form
\begin{equation}
\frac{1}{\tau}\bigl(\tau T_0^0\bigr)^{\cdot} - \frac{\dot a}{a} T_1^1
-\frac{\dot b}{b} T_2^2  - \frac{\dot c}{c} T_3^3 = 0.
\label{emcon}
\end{equation}
This equation can be rewritten as
\begin{equation}
\dot{\tilde{T}}_0^0 = \frac{\dot \tau}{\tau}\Bigl(\tilde{T}_{1}^{1} - \tilde{T}_{0}^{0}\Bigr)
+ 2 \eta \Bigl(\frac{\dot a^2}{a^2} + \frac{\dot b^2}{b^2} + \frac{\dot c^2}{c^2}\Bigr).
\label{Biden}
\end{equation}
Recall that \eqref{inv} gives
$$(m -{\cD}) \dot{S}_0 - {\cG} \dot{P}_0 = 0.$$
In view of that after a little manipulation from \eqref{Biden} we obtain
\begin{equation}
{\dot \ve} + \frac{\dot \tau}{\tau} \omega - (\xi + \frac{4}{3}
\eta) \frac{{\dot \tau}^2}{\tau^2} + 4 \eta (\kappa T_0^0 +
\Lambda) = 0, \label{vep}
\end{equation}
where
\begin{equation}
\omega = \ve + p,
\label{thermal}
\end{equation}
is the thermal function. For further purpose we would like to note that
in absence of shear viscosity from Eqs. \eqref{a4} and \eqref{Biden} one finds
\begin{equation}
\kappa \tilde{T}_0^0 = 3 H^2 - \Lambda + C_{00}, \quad C_{00} =
{\rm const.} \label{veHrel}
\end{equation}
where in analogy with Hubble constant introduce the quantity $H$, such
that
\begin{equation}
\frac{\dot {\tau}}{\tau} = \frac{\dot {a}}{a}+\frac{\dot {b}}{b} +
\frac{\dot {c}}{c} = 3 H.
\label{hubc}
\end{equation}
Then \eqref{dtau1} and \eqref{vep} in account of \eqref{total} can be
rewritten as
\begin{subequations}
\label{HVe}
\begin{eqnarray}
\dot {H} &=& \frac{\kappa}{2}\bigl(3 \xi H - \omega\bigr) -
\bigl(3 H^2 - \kappa \ve - - \Lambda\bigr) + \frac{\kappa}{2}
\bigl(m S + {\cD} S + {\cG} P - 2 \lambda F \bigr),  \label{H}\\
\dot {\ve} &=& 3 H\bigl(3 \xi H - \omega\bigr) + 4 \eta \bigl(3
H^2 - \kappa \ve  - \Lambda\bigr) - 4 \eta \kappa \bigl( m S -
\lambda F\bigr). \label{Ve}
\end{eqnarray}
\end{subequations}
Thus, the metric functions are found explicitly in terms of $\tau$
and viscosity. To write $\tau$ and components of spinor field as
well and scalar one we have to specify $F$ in ${\cal L}_{\rm sp}$.
In the next section we explicitly solve Eqs. \eqref{V} and
\eqref{HVe} for some concrete value of $F$.

The Eqs. \eqref{HVe} can be written in terms of dynamical scalar
as well. For this purpose let us introduce the dynamical scalars
such as the expansion and the shear scalar as usual
\begin{equation}
\theta = u^\mu_{;\mu},\quad
\sigma^2 = \frac{1}{2} \sigma_{\mu\nu} \sigma^{\mu\nu},
\label{dynscal}
\end{equation}
where
\begin{equation}
\sigma_{\mu\nu} = \frac{1}{2} \Bigl(u_{\mu;\alpha} P^{\alpha}_{\nu}
+ u_{\nu;\alpha} P^{\alpha}_{\mu}\Bigr) - \frac{1}{3} \theta P_{\mu \nu}.
\label{shearten}
\end{equation}
Here $P$ is the projection operator obeying
\begin{equation}
P^2 = P.
\end{equation}
For the space-time with signature $(+, \,-,\,-,\,-)$ it has the form
\begin{equation}
P_{\mu\nu} = g_{\mu\nu} - u_\mu u_\nu, \quad
P^\mu_\nu = \delta^\mu_\nu - u^\mu u_\nu.
\label{projec}
\end{equation}
For the BI metric the dynamical scalar has the form
\begin{equation}
\theta = \frac{\dot {a}}{a}+\frac{\dot {b}}{b} +
\frac{\dot {c}}{c} = \frac{\dot {\tau}}{\tau},
\label{expan}
\end{equation}
and
\begin{equation}
2 \sigma^2 = \frac{\dot {a}^2}{a^2}+\frac{\dot {b}^2}{b^2} +
\frac{\dot {c}^2}{c^2} - \frac{1}{3} \theta^2.
\label{shesc}
\end{equation}
In account of \eqref{abc} one can also rewrite share scalar as
\begin{equation}
2 \sigma^2 = \frac{6 (X_1^2 + X_1 X_3 + X_3^2)}{9 \tau^2} e^{-4
\kappa \int \eta dt}.
\end{equation}
From \eqref{00} now yields
\begin{equation}
\frac{1}{3} \theta^2 - \sigma^2 = \kappa \Bigl[
 mS -\lambda F + \ve\Bigr] + \Lambda
\label{sh00}
\end{equation}
The Eqs. \eqref{HVe} now can be written in terms of $\theta$ and
$\sigma$ as follows
\begin{subequations}
\label{TS}
\begin{eqnarray}
\dot {\theta} &=& \frac{3\kappa}{2}\bigl(\xi \theta - \omega\bigr) -
 \frac{3\kappa}{2} \bigl(
m S - {\cD} S - {\cG} P \bigr) -
3 \sigma^2,  \label{theta}\\
\dot {\ve} &=& \theta \bigl(\xi \theta - \omega\bigr) + 4 \eta \sigma^2.
\label{Vsig}
\end{eqnarray}
\end{subequations}
Note that the Eqs. \eqref{TS} without spinor and scalar field
contributions coincide with the ones given in \cite{baner}.

        \section{Some special solutions}

In this section we first solve the spinor field equations for some special
choice of $F$, which will be given in terms of $\tau$. Thereafter, we
will study the system \eqref{HVe} in details and give explicit solution
for some special cases.

    \subsection{Solutions to the spinor field equations}
As one sees, introduction of viscous fluid has no direct effect on
the system of spinor field equations \eqref{V}. Viscous fluid has an
implicit influence on the system through $\tau$. A detailed analysis
of the system in question can be found in \cite{sahaprd}. Here
we just write the final results.

    \subsubsection{Case with $F = F(I)$}

Here we consider the case when the nonlinear spinor field
is given by
$F =  F(I).$
As in the case with minimal coupling from \eqref{S0} one finds
\begin{equation}
S = \frac{C_0}{\tau}, \quad C_0 = {\rm const.}
\label{stau}
\end{equation}
For  components of spinor field we find~\cite{sahaprd}
\begin{eqnarray}
\psi_1(t) &=& \frac{C_1}{\sqrt{\tau}} e^{-i\beta}, \quad
\psi_2(t) = \frac{C_2}{\sqrt{\tau}} e^{-i\beta},  \nonumber\\
\label{spef}\\
\psi_3(t) &=& \frac{C_3}{\sqrt{\tau}} e^{i\beta}, \quad
\psi_4(t) = \frac{C_4}{\sqrt{\tau}} e^{i\beta},
\nonumber
\end{eqnarray}
with $C_i$ being the integration constants and
are related to $C_0$ as
$C_0 = C_{1}^{2} + C_{2}^{2} - C_{3}^{2} - C_{4}^{2}.$ Here
$\beta = \int(m - {\cD}) dt$.

For the components of the spin current from \eqref{spincur} we find
\begin{eqnarray}
j^0 &=& \frac{1}{\tau} \bigl[C_{1}^{2} + C_{2}^{2} + C_{3}^{2} +
C_{4}^{2}\bigr],\quad j^1 = \frac{2}{a\tau} \bigl[C_{1} C_{4} +
C_{2} C_{3}\bigr] \cos (2\beta),
\nonumber \\
j^2 &=& \frac{2}{b\tau} \bigl[C_{1} C_{4} - C_{2} C_{3}\bigr] \sin
(2\beta),\quad j^3 = \frac{2}{c\tau} \bigl[C_{1} C_{3} - C_{2}
C_{4}\bigr] \cos (2\beta), \nonumber
\end{eqnarray}
whereas, for the projection of spin vectors on the $X$, $Y$ and $Z$
axis we find
\begin{eqnarray}
S^{23,0} = \frac{C_1 C_2 + C_3 C_4}{b c\tau},\quad
S^{31,0} = 0,\quad
S^{12,0} = \frac{C_1^2 - C_2^2 + C_3^2 - C_4^2}{2ab\tau}. \nonumber
\end{eqnarray}
The total charge of the system in a volume $\cal{V}$ in this case is
\begin{equation}
Q = [C_1^2 + C_{2}^{2} + C_{3}^{2} + C_{4}^{2}] \cal{V}.
\end{equation}
Thus, for $\tau \ne 0$ the components of spin current and
the projection of spin vectors are singularity-free and the total
charge of the system in a finite volume is always finite.
Note that, setting $\lambda = 0$, i.e., $\beta = m t$ in the
foregoing expressions one get the results for the linear spinor field.

    \subsubsection{Case with $F = F(J)$}

Here we consider the case with
$F = F(J).$
In this case we assume the spinor field to be massless.
Note that, in the unified
nonlinear spinor theory of Heisenberg, the massive term remains
absent, and according to Heisenberg, the particle mass should be
obtained as a result of quantization of spinor prematter~
\cite{massless}. In the nonlinear generalization of classical field
equations, the massive term does not possess the significance that
it possesses in the linear one, as it by no means defines total
energy (or mass) of the nonlinear field system. Thus without losing
the generality we can consider massless spinor field putting $m\,=\,0.$
Then from \eqref{P0} one gets
\begin{equation}
P = D_0/\tau, \quad D_0 = {\rm const.}
\label{ptau}
\end{equation}
In this case the spinor field components take the form
\begin{eqnarray}
\psi_1 &=&\frac{1}{\sqrt{\tau}} \bigl(D_1 e^{i \sigma} +
iD_3 e^{-i\sigma}\bigr), \quad
\psi_2 =\frac{1}{\sqrt{\tau}} \bigl(D_2 e^{i \sigma} +
iD_4 e^{-i\sigma}\bigr), \nonumber \\
\label{J}\\
\psi_3 &=&\frac{1}{\sqrt{\tau}} \bigl(iD_1 e^{i \sigma} +
D_3 e^{-i \sigma}\bigr),\quad
\psi_4 =\frac{1}{\sqrt{\tau}} \bigl(iD_2 e^{i \sigma} +
D_4 e^{-i\sigma}\bigr). \nonumber
\end{eqnarray}
The integration constants $D_i$
are connected to $D_0$ by
$D_0=2\,(D_{1}^{2} + D_{2}^{2} - D_{3}^{2} -D_{4}^{2}).$
Here we set $\sigma = \int {\cG} dt$.

For the components of the spin current from \eqref{spincur} we find
\begin{eqnarray}
j^0 &=& \frac{2}{\tau}
\bigl[D_{1}^{2} + D_{2}^{2} + D_{3}^{2} + D_{4}^{2}\bigr],\quad
j^1 = \frac{4}{a\tau}
\bigl[D_{2} D_{3} + D_{1} D_{4}\bigr] \cos (2 \sigma), \nonumber\\
j^2 &=& \frac{4}{b\tau} \bigl[D_{2} D_{3} - D_{1} D_{4}\bigr] \sin
(2 \sigma),\quad j^3 = \frac{4}{c\tau} \bigl[D_{1} D_{3} - D_{2}
D_{4}\bigr] \cos (2 \sigma), \nonumber
\end{eqnarray}
whereas, for the projection of spin vectors on the $X$, $Y$ and $Z$
axis we find
\begin{eqnarray}
S^{23,0} = \frac{2(D_{1} D_{2} + D_{3} D_{4})}{b c\tau},\quad
S^{31,0} = 0,\quad
S^{12,0} = \frac{D_{1}^{2} - D_{2}^{2} + D_{3}^{2} - D_{4}^{2}}{2ab\tau}
\nonumber
\end{eqnarray}
We see that for any nontrivial $\tau$ as in previous case
the components of spin current and
the projection of spin vectors are singularity-free and the total
charge of the system in a finite volume is always finite.

 \subsection{Determination of $\tau$}

In this subsection we simultaneously solve the system of equations
for $\tau$ and $\ve$. Since setting $m = 0$ in the equations for
$F= F(I)$ one comes to the case when $F = F(J)$, we consider the
case with $F$ being the function of $I$ only. Let $F$ be the power
function of $S$, i.e., $F = S^n$. As it was established earlier,
in this case $S = C_0/\tau$, or setting $C_0 = 1$ simply $S =
1/\tau$. Evaluating ${\cD}$ in terms of $\tau$ we then come to the
following system of equations
\begin{subequations}
\label{tauve}
\begin{eqnarray}
{\ddot \tau} &=&  \frac{3\kappa}{2} \xi {\dot \tau} +
\frac{3\kappa}{2} \Bigl(\frac{m}{\tau} +\frac{\lambda
(n-2)}{\tau^{n}}
 +  \ve - p
\Bigr) \tau + 3 \Lambda \tau,\label{dtaun}\\
{\dot \ve} &=& - \frac{\dot \tau}{\tau} \omega + (\xi +
\frac{4}{3} \eta) \frac{{\dot \tau}^2}{\tau^2} - 4 \eta
\Bigl[\kappa \Bigl(\frac{m}{\tau} - \frac{\lambda}{\tau^n}\Bigr) +
\Lambda\Bigr], \label{vepn}
\end{eqnarray}
\end{subequations}
or in terms of $H$
\begin{subequations}
\label{HVen}
\begin{eqnarray}
\dot \tau &=& 3 H \tau, \label{tauH}\\
\dot {H} &=& \frac{1}{2}\bigl(3 \xi H - \omega\bigr) - \bigl(3 H^2
- \kappa\ve  - \Lambda\bigr) + \frac{\kappa}{2} \Bigl(
\frac{m}{\tau} + \frac{\lambda (n-2)}{\tau^{n}} \Bigr),
\label{Hn}\\
\dot {\ve} &=& 3 H\bigl(3 \xi H - \omega\bigr) + 4 \eta \bigl(3
H^2 - \kappa \ve  - \Lambda\bigr) - 4 \eta \kappa\Bigl[
\frac{m}{\tau} - \frac{\lambda}{\tau^n}\Bigr]. \label{Ven}
\end{eqnarray}
\end{subequations}
Here $\eta$ and $\xi$ are the bulk and shear viscosity, respectively and
they are both positively definite, i.e.,
\begin{equation}
\eta > 0, \quad \xi > 0.
\end{equation}
They may be either constant or function of time or energy. We consider
the case when
\begin{equation}
\eta = A \ve^{\alpha}, \quad \xi = B \ve^{\beta},
\label{etaxi}
\end{equation}
with $A$ and $B$ being some positive quantities.
For $p$ we set as in perfect fluid,
\begin{equation}
p = \zeta \ve, \quad \zeta \in (0, 1].
\label{pzeta}
\end{equation}
Note that in this case $\zeta \ne 0$, since for dust pressure, hence
temperature is zero, that results in vanishing viscosity.

The system \eqref{HVen} without spinor field have been extensively
studied in literature either partially \cite{murphy,huang,baner}
or as a whole \cite{belin}. Here we try to solve the system
\eqref{tauve} for some particular choice of parameters.

        \subsubsection{Case with bulk viscosity}

Let us first consider the case with bulk viscosity alone setting coefficient of
shear viscosity $\eta = 0$. We also demand the coefficient of bulk viscosity be
inverse proportional to expansion, i.e.,
\begin{equation}
\xi \theta = 3 \xi H = C_2, \quad C_2 = {\rm const.}
\label{bvx}
\end{equation}
Inserting $\eta = 0$, \eqref{bvx} and \eqref{pzeta} into \eqref{Ven} one finds
\begin{equation}
\ve = \frac{1}{1+\zeta} \Bigl[C_2 -
\frac{C_3}{\tau^{1+\zeta}}\Bigr], \qquad C_3 = {\rm
const.}\label{vecase1}
\end{equation}
Then from \eqref{dtaun} we get the following equation for determining $\tau$:
\begin{equation}
\ddot \tau = \frac{3\kappa}{2} m + 3\Bigl[\frac{C_2}{2}\kappa +
\Lambda\Bigr] \tau + \frac{3\kappa(1 - \zeta)}{2(1+\zeta)}
\frac{C_2 \tau^{1+\zeta} - C_3}{\tau^\zeta} +
\frac{3\kappa}{2}\frac{\lambda (n-2)}{\tau^{n-1}} \equiv {\cal
F}(q,\tau), \label{dettau1}
\end{equation}
where $q$ is the set of problem parameters. As one sees, the right hand side of the Eq. \eqref{dettau1}
is a function of $\tau$, hence can be solved in quadrature \cite{kamke}.
We solve the Eq. \eqref{dettau1} numerically. It can be noted that the Eq. \eqref{dettau1}
can be viewed as one describing the motion of a single particle. Sometimes it is useful to plot
the potential of the corresponding equation which in this case is
\begin{equation}
{\cal U}(q,\tau) = - 2 \int {\cal F}(q,\tau) d\tau. \label{poten1}
\end{equation}
The problem parameters are chosen as follows: $\kappa = 1$, $m =
1$, $\lambda = 0.5$, $\zeta = 1/3$, $n = 4$, $C_2 = 2$ and $C_3 =
1$. Here we consider the cases with different $\Lambda$, namely
with $\Lambda = -2, 0, 1$, respectively. The initial value of
$\tau$ is taken to be a small one, whereas, the first derivative
of $\tau$, i.e., $\dot \tau$ at that point of time is calculated
from \eqref{veHrel}. In Fig. \ref{rfsppoten} we have illustrated
the potential corresponding to Eq. \eqref{dettau1}. As one sees,
independent to the sign of $\Lambda$ we have the expanding mode of
evolution, though a positive $\Lambda$ accelerates the process,
while the negative one decelerates. Corresponding behavior of
$\tau$ is given in Fig. \ref{rfsptauall}.

\myfigures{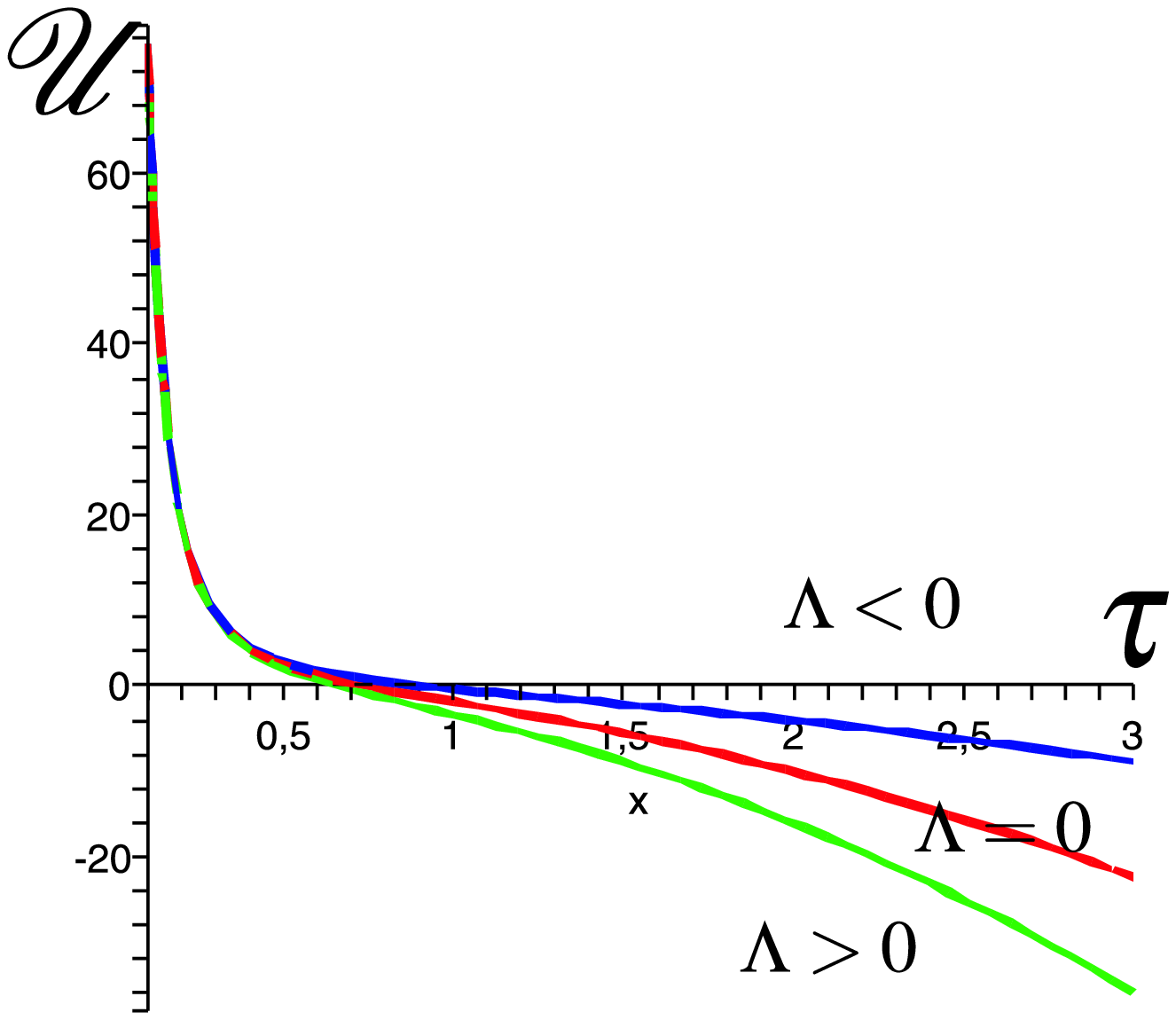}{0.45}{View of the potential corresponding to
the different sign of the $\Lambda$
term.}{0.45}{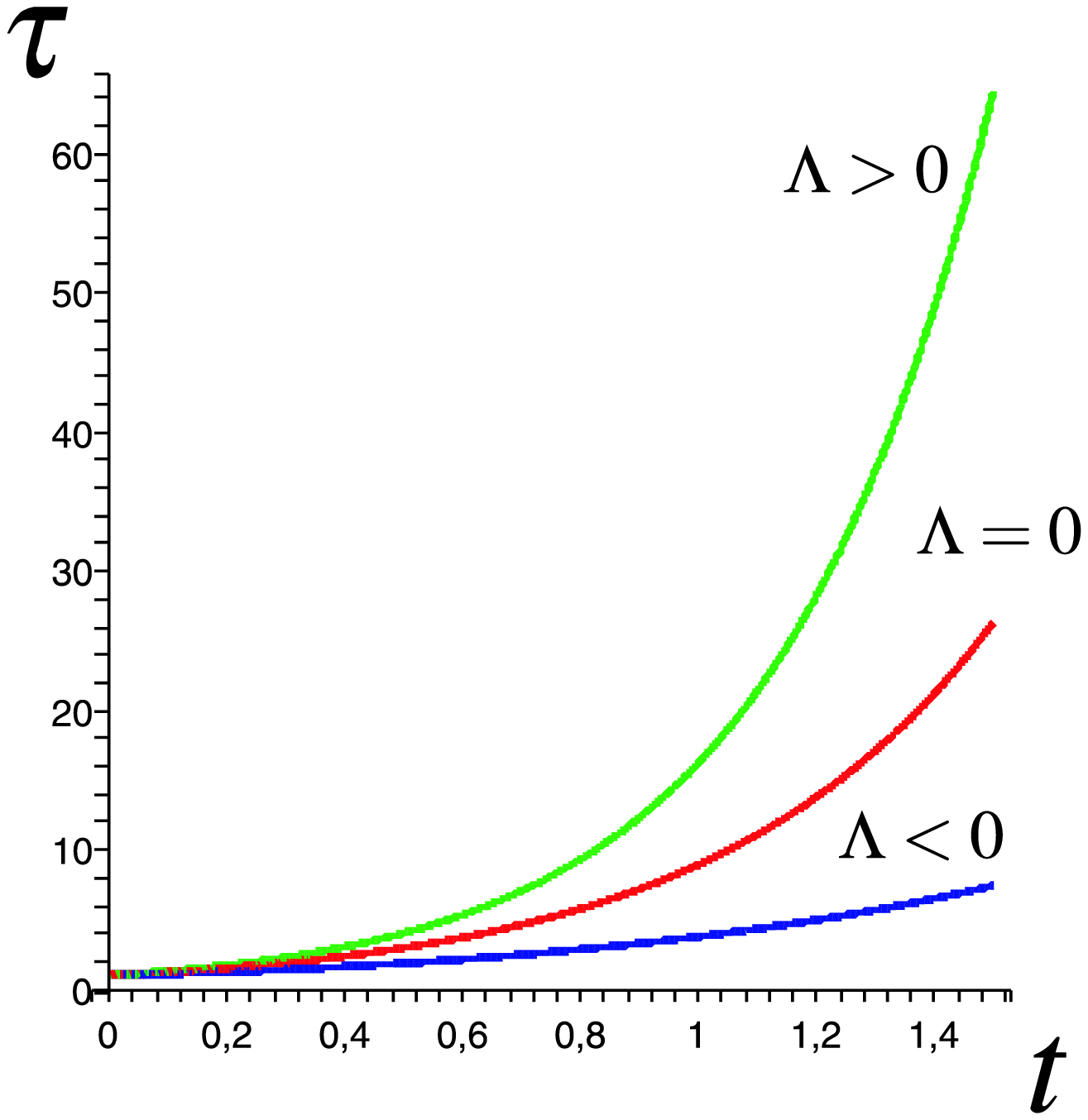}{0.45}{Evolution of $\tau$ depending on
the signs of the $\Lambda$ term.}{0.45}

        \subsubsection{Case with bulk and shear viscosities}
Let us consider a more general case. Following \cite{Visprd04} we
choose the shear viscosity being proportional to the expansion,
namely,
\begin{equation}
\eta = - \frac{3}{2\kappa} H = -\frac{1}{2\kappa} \theta.
\label{sbv}
\end{equation}
In absence of spinor field this assumption leads to
\begin{equation}
3H^2 = \kappa \ve + C_4, \quad C_4 = {\rm const.} \label{sbvrel}
\end{equation}
It can be shown that the relation \eqref{sbvrel} in our case can
be achieved only for massless spinor field with the nonlinear term
being
$$ F = F_0 S^{2(\kappa - 1)/\kappa}.$$
Equation for $\tau$ in this case has the form
\begin{equation}
\tau \ddot \tau - 0.5 (1 - \zeta) \dot \tau^2 - 1.5 \kappa \xi
\tau \dot \tau - 3[\Lambda - 0.5(1 - \zeta)C_4 - \lambda F_0
\tau^{2(1-\kappa)/\kappa}] \tau^2 = 0. \label{sbvtau}
\end{equation}
In case of $\xi = {\rm const.}$ and $\lambda = 0$ there exists
several special solutions available in handbooks on differential
equations. But for nonzero $\lambda$ we can investigate this
equation only numerically. We consider the case when the bulk
viscosity is given by a constant. Taking this into account for
problem parameters we set $\zeta = 1/3$, $\xi = 1$, $F_0 = 1$,
$\lambda = 0.5$ and $C_4 = 1$. We study the role of $\Lambda$
term. In doing this we consider the cases with positive, negative
and trivial $\Lambda$. Since the nonlinear term in this case
depends of $\kappa$, we also consider the cases with different
$\kappa$, namely with $\kappa > 1$ and $\kappa < 1$. In Figs.
\ref{rfspk05all} and \ref{rfspk15all} the evolution of $\tau$ is
illustrated for $\kappa < 1$ and $\kappa > 1$, respectively. In
case of $\kappa < 1$ we have non-periodic mode of evolution for
all $\Lambda$, while for $\kappa > 1$ a negative $\Lambda$ gives a
non-periodic mode of expansion. A non-negative $\Lambda$ in this
case gives an ever expanding mode of evolution.

\myfigures{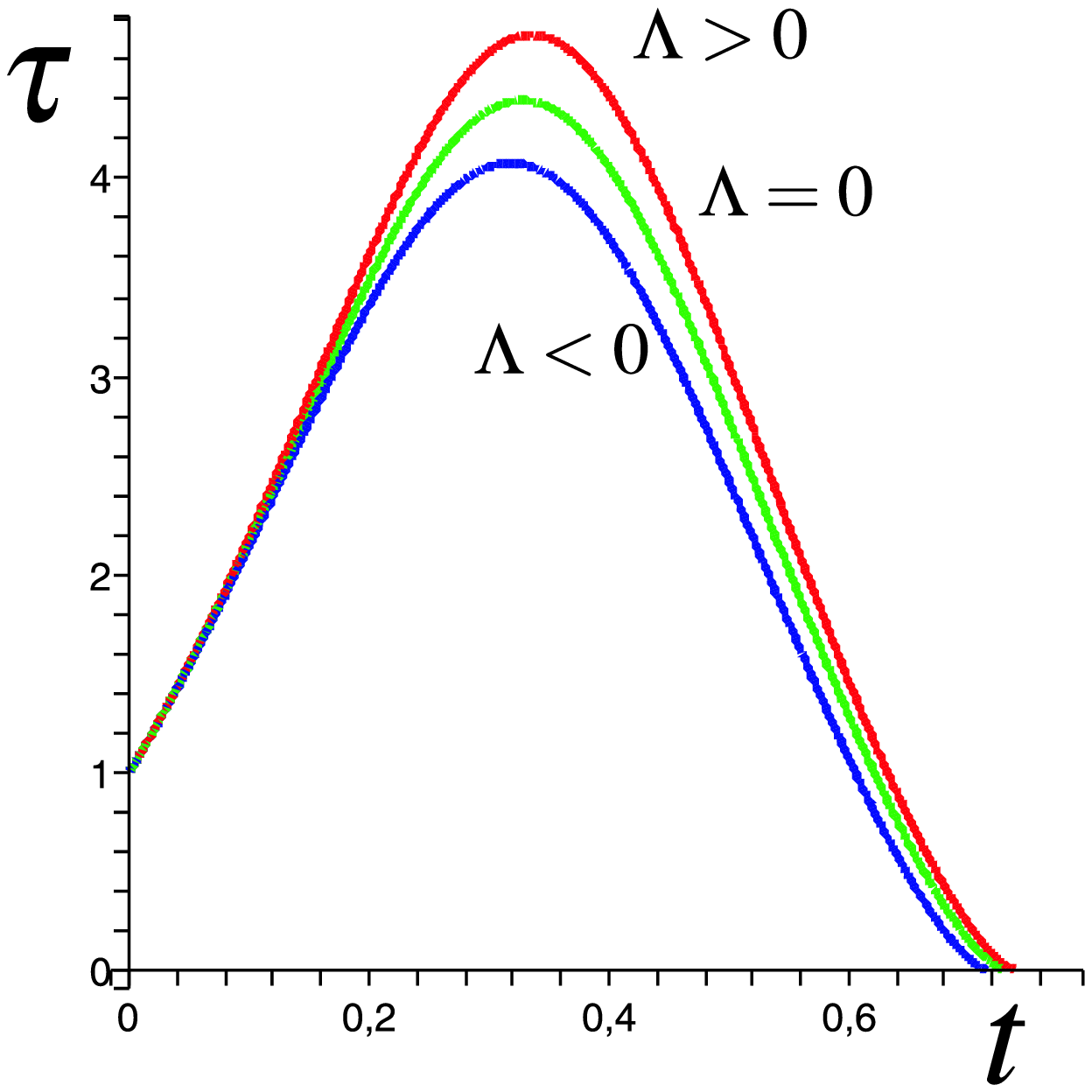}{0.45}{Evolution of the universe with
nontrivial $\Lambda$ term and $\kappa <
1$.}{0.45}{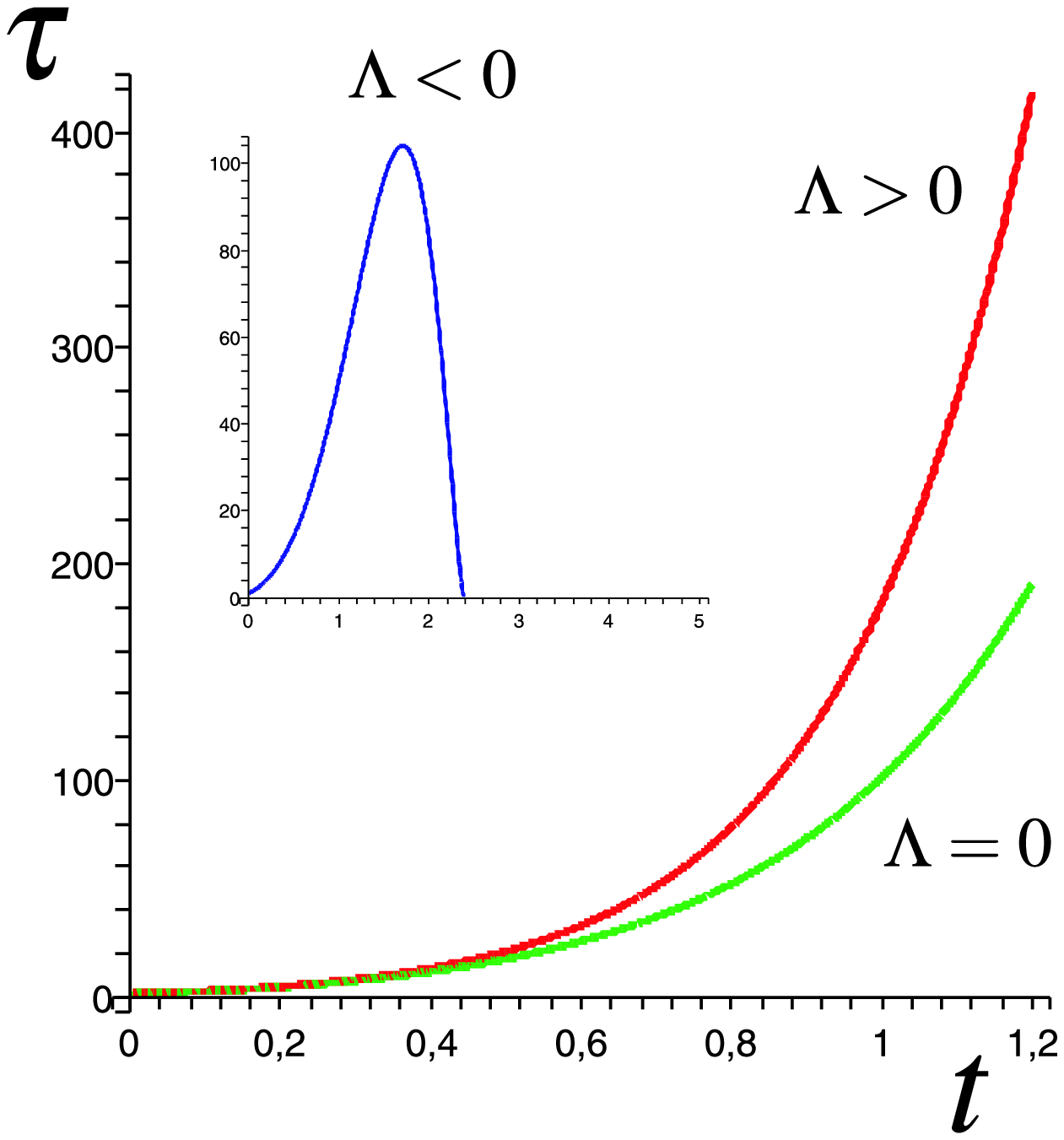}{0.45}{Evolution of the universe for
different values of $\Lambda$ term with $\kappa > 1$. }{0.45}


               \section{Conclusion}
We consider a self consistent system of nonlinear spinor and
gravitational fields within the framework of Bianchi type-I
cosmological model filled with viscous fluid. The spinor filed
nonlinearity is taken to be some power law of the invariants of
bilinear spinor forms. Solutions to the corresponding equations
are given in terms of the volume scale of the BI space-time, i.e.,
in terms of $\tau = a b c$. The system of equations for
determining $\tau$, energy-density of the viscous fluid $\ve$ and
Hubble parameter $H$ has been worked out. Exact solution to the
aforementioned system has been given only for the case of bulk
viscosity. As one sees from \eqref{HVe} or \eqref{TS}, the system
in question is a multi-parametric one and may have several
solutions depending on the choice of the problem parameters. As
one sees, solutions can be non-periodic independent to the sign of
$\Lambda$ term. Given the richness of \eqref{tauve} we plan to
give qualitative analysis of this system in near future.


\newcommand{\hnl}{\htmladdnormallink}

\end{document}